\documentclass[10pt]{article}
\usepackage[utf8]{inputenc}
\usepackage{authblk}

\usepackage{graphicx}
\usepackage[cmex10]{amsmath}
\usepackage{amssymb,amsfonts}
\usepackage{algorithmic}
\usepackage[ruled,vlined]{algorithm2e}
\usepackage{listings}
\usepackage{textcomp}

\usepackage[backend=bibtex,sorting=none,style=ieee]{biblatex}
\begin{document}
\title{Improving Bitcoin Transaction Propagation \\by Leveraging Unreachable Nodes}

\author[1]{Federico Franzoni\thanks{This author is partly supported by the Spanish Ministry of Economy and Competitiveness under the Maria de Maeztu Units of Excellence Programme (MDM-2015-0502).}}
\author[1]{Vanesa Daza\thanks{This author was supported by Project RTI2018-102112-B-I00 (AEI/FEDER,UE).}} %
\affil[1]{Universitat Pompeu Fabra, Barcelona, Spain\\\texttt{\{federico.franzoni,vanesa.daza\}@upf.edu}}
\date{}
\maketitle

\begin{abstract}
The Bitcoin P2P network is at the core of all communications between clients.
The reachable part of this network has been explored and analyzed by numerous studies.
Unreachable nodes, however, are, in most part, overlooked.
Nonetheless, they are a relevant part of the network and play an essential role in the propagation of messages.
In this paper, we focus on transaction propagation and show that increasing the participation of unreachable nodes can potentially improve the robustness and efficiency of the network.
In order to do that, we propose a few changes to the network protocol.
Additionally, we design a novel transaction propagation protocol that explicitly involves unreachable nodes to provide better protection against deanonymization attacks.
Our solutions are simple to implement and can effectively bring immediate benefits to the Bitcoin network.
\end{abstract}

\section{Introduction}
The Bitcoin P2P network is the channel through which clients exchange transactions and blocks.
The characteristics and behavior of the nodes in this network have been extensively studied in the literature, with particular attention to efficiency and security.
However, most papers only focus on the reachable portion of the network, leaving unreachable nodes out of their scope~\cite{donet2014bitcoin,miller2015discovering,mariem2020glitters}.
Nevertheless, these nodes represent more than 90\% of the whole network~\cite{wang2017towards,delgado2018txprobe}.
The reason for this imbalance resides in the inability of measuring tools to create connections towards unreachable nodes, which, by definition, only establish outgoing connections.
As a result, the unreachable part of the network is often overlooked, and its relevance is underestimated.

In this paper, we analyze the importance of unreachable nodes in the propagation of messages and show how their participation can be beneficial to the network.
In particular, we study the characteristics of these nodes, as emerged from state-of-the-art research, and identify some of their strengths and weaknesses compared to reachable nodes.
We then propose changes to the protocol to improve the connectivity of the network as well as the efficiency of message propagation.

Additionally, we show that unreachable nodes are protected from adversaries that connect to the victims.
Based on this characteristic, we design a novel transaction propagation protocol that potentially improves security against deanonymization attacks.
Our solution explicitly involves unreachable nodes in the propagation pattern and exploits their position in the network to conceal the source of the message.
We thoroughly justify our design choices and study the security of our protocol against an eavesdropper adversary~\cite{fanti2017deanonimization}.

Our contribution includes:
\begin{itemize}
	\item we study the characteristics of unreachable nodes based on state-of-the-art research, and show have natural protection against a number of attacks;
	\item we show how unreachable nodes can play an important role in the network and propose changes to the protocol in order to achieve more robustness and improve the efficiency of message propagation;
	\item we design a new transaction propagation protocol aimed at improving anonymity and study its resilience against powerful adversaries.
\end{itemize}

\section{Background}
Unreachable nodes are typically associated with NATs.
In fact, Carrier-Grade NATs (CG-NATs) are the primary cause for the unreachability of a node, as most ISPs use this technology to grant access to a larger number of devices.
In this section, we provide more details about NAT and describe the current Bitcoin transactions propagation protocol, which we will modify to provide better anonymity.

\subsection{NAT and P2P networks}
\textit{Network Address Translation} (NAT) \cite{rfc1999nat} is a method to map IP addresses between incompatible networks.
The most common type, known as \textit{Network Address and Port Translation} (NAPT), is often used to connect private networks to the Internet without the need to assign a unique address to each device.
NAPT is often regarded as a solution to the IPv4 address exhaustion problem \cite{richter2015primer}, since it allows a large number of devices to connect through a shared IP address.
As a side effect, such devices cannot be reached from the Internet, unless they first open a connection.

While this is not a problem in a client-server setting, it is a serious limitation for P2P networks.
Notably, it prevents NATted nodes from connecting to each other.
To overcome this limitation, \textit{NAT traversal} techniques have been devised \cite{hu2005nat}.

The Bitcoin reference client implements \textit{Universal Plug-and-Play} (UPnP), which, however, is incompatible with CG-NATs, as it needs direct access from the host.
Furthermore, the UPnP option is disabled by default due to a known vulnerability in the protocol.
As a consequence, NATted nodes only establish outbound connections, which in the reference client are limited to just 8.

\subsection{Transaction Propagation and Anonymity}
Forwarding a transaction in Bitcoin is a three-step process.
First, the sender transmits an inventory (INV) message to advertise the hash of the transaction.
The receiver then checks the hash and, if it is unknown, requests the full transaction data with a GETDATA message.
Finally, the sender transmits the full transaction in a TX message.
The INV-based transmission allows sending the transaction only to those nodes which still have not received it.
When a node receives a new transaction, it relays it to its peers following the same process.

The relay step is fundamental in determining the propagation pattern.
Since 2015, Bitcoin adopts the \textit{diffusion spreading} protocol, where nodes relay transactions to each neighbor with an independent, exponential delay.
Newly-generated transactions are transmitted in the same way by their source.

As shown in \cite{fanti2017deanonimization}, the pattern generated by this gossip-like protocol leads to possible deanonymization attacks based on the so-called \textit{rumor centrality}~\cite{shah2012rumor}.
In simple words, since a transaction spreads symmetrically from each node to its peers, it is possible to determine the origin of the spreading (i.e., the first node that transmitted the transaction) by observing the state of its propagation through the network.
Given that the source broadcasts the transaction in the same way as the relays, detecting the origin of the propagation often means identifying the creator of the transaction (in terms of the device address).

Based on this fact, several attacks~\cite{koshy2014analysis,fanti2017deanonimization,biryukov2019deanonymization} have been shown where an \textit{eavesdropper adversary} connects to all reachable nodes and applies the so-called \textit{first-spy} estimator, which simply associates a transaction to the first node that relays it.
Fanti et al.~\cite{fanti2017anonymity} showed that this type of strategy often has very high levels of accuracy.

\section{Bitcoin Unreachable Nodes}
\label{sect:unreachable}
Nodes in P2P networks are traditionally divided in research into reachable and unreachable.
A node is called \textit{reachable} if it can accept incoming connections from other peers.
Otherwise, it is called \textit{unreachable}.

Nodes can be unreachable because they are protected by a firewall, connecting through a proxy or, most typically, being hosted in a private network, behind a NAT device. 
Less commonly, nodes purposely choose not to accept incoming connections.

Reachable and unreachable nodes are often named servers and clients, respectively, to recall the ability of the firsts to accept connections and the fact that the seconds connect to them.
However, there is no client-server relationship among them, as they follow the same P2P protocol.
A more precise classification commonly used in other P2P-related papers, distinguish between \textit{routable} and \textit{non-routable} peers, and calls \textit{unreachable} those peers that are known to other peers but cannot be contacted (e.g., because they are offline or only accept connections from known peers)~\cite{essaid2020bitcoin}.
In the following, the terms reachable and unreachable will be used to indicate routable and non-routable.
Furthermore, for the sake of simplicity, reachable and unreachable nodes will be denoted by \textit{R nodes} and \textit{U nodes}, respectively.

Despite their relevance, U nodes have been marginally covered by state-of-the-art research.
Most Bitcoin network-related analyses focus on R nodes \cite{decker2013information, donet2014bitcoin, feld2014analyzing, miller2015discovering}, leaving U nodes out of scope.
The statistics given by these works hardly give a precise account of U nodes, as they do not distinguish between offline nodes and nodes that are actually out of reach.
However, virtually all of them show that the number of U nodes is much greater than R nodes (estimates go from 10~\cite{delgado2018txprobe} to 30 times more~\cite{wang2017towards}).
Furthermore, studies showed that regular Bitcoin users tend to use U nodes, while R nodes are mostly run in data centers~\cite{neudecker2019characterization,wang2017towards}.

As noted in \cite{neudecker2019network}, U nodes can contribute to the robustness of the network, as they increase connectivity and they are harder to attack for adversaries without access to core infrastructure.
One of the goals of this paper will be to improving the efficiency and robustness of data propagation by increasing the participation of U nodes in the network.

U nodes are often overlooked due to the difficulty of connecting measurement tools to all of them at the same time, which is commonly done with R nodes. 
The only way to study U nodes is to deploy R nodes and wait for U nodes to connect to them.
By adopting this approach, Wang et al. \cite{wang2017towards} were able to study U nodes in detail.
As of today, their work is the only known global analysis of U nodes.
Besides it, the only papers focusing on U nodes are those targeting them for deanonymization~\cite{biryukov2014deanonymisation,biryukov2015tor,mastan2018new}.
The interest in this kind of attacks stems from the relative difficulty of targeting U nodes at a global level.
In fact, like U nodes are hard to study, they are also hard to target by an attacker, which has to deal with the inability to open connections to them.
Specifically, U nodes are hard to include in observation-based attacks \cite{fanti2017deanonimization} and unsolicited-message-based attacks \cite{rossow2013p2pwned}.
Similarly, U nodes are also immune to many network-level attacks, such as eclipse attacks~\cite{heilman2015eclipse}, topology-inferring attacks~\cite{miller2015discovering}, and partitioning attacks~\cite{tran2020stealthier}.

NATted nodes are also hard to distinguish, since they can share the same IP address.
This is why network-wide deanonymization attacks against U nodes usually require fingerprinting techniques for identifying nodes individually~\cite{biryukov2014deanonymisation,mastan2018new}.
Nonetheless, U nodes are very susceptible to deanonymization attacks in case they directly connect to the adversary.
In this case, even a simple first-spy estimator can obtain a very high level of accuracy~\cite{wang2017towards}.
In this paper, we propose a new propagation protocol that allows protecting both R and U nodes from deanonymization.

\section{Our Solution}
We propose some changes to the network protocol, which leverage the specificity of unreachable nodes to improve the efficiency and security of the network.
In particular, we propose the following changes:
\begin{itemize}
	\item Explicitly distinguish reachable and unreachable nodes;
	\item Increase connections from unreachable nodes;
	\item Disable advertisement of unreachable addresses;
	\item Adopt the propagation protocol described in \S \ref{protocol}.
\end{itemize}

\subsection{Network Changes}
We first describe the changes to basic network protocol behavior that allow us to improve security and efficiency. 

\subsubsection{Explicitly distinguish between R and U nodes}
Although there is some difference in the behavior of R nodes and U nodes in the reference client, the Bitcoin network protocol does not make any explicit distinction between them.
However, our solution is based on the different characteristics shown by the two types of node, as shown in  \S \ref{sect:unreachable}.
As such, explicitly distinguish between R nodes and U nodes is a necessary step.

Different strategies can be followed by a node to determine its reachability.
A naive approach would be to verify if the client accepts incoming connections.
However, it might be the case that a node is accepting connections but its address is unreachable from the outside.
A better approach is to have the node connect to its own address, as seen by its peers, and set itself reachable, if the attempt succeeds, and unreachable, otherwise.

\subsubsection{Increase U nodes connections}
The second modification we propose is to increase the number of outbound connections of U nodes.
This change has several effects.
Firstly, it helps leveling the imbalance of connectivity between R and U nodes.
In fact, while R nodes can reach 125 connections, U nodes only maintain up to 8, corresponding to their outbound peers.
On the other hand, inbound slots are often underutilized by R nodes~\cite{decker2013information}, which means they can handle a higher number of connections.

Secondly, increasing the number of peers means receiving, and relaying, more transactions per amount of time.
Given the great number of U nodes, even a small increase in their connections might produce a significant improvement in the propagation speed of transactions and blocks.

Furthermore, from a security perspective, it has been shown how increasing the number of outbound connections can improve resistance against DoS attacks~\cite{neudecker2019network}, eclipse attacks~~\cite{heilman2015eclipse}, and isolation attacks~\cite{apostolaki2017hijacking}.

Finally, a higher number of connections for U nodes can be beneficial for the anonymity of our propagation protocol, as we will show in \S \ref{protocol}.

\subsubsection{Do not advertise U nodes addresses}
In the current protocol, U nodes, like R nodes, advertise their public address to their peers.
These addresses represent 90\% of those being spread through the network~\cite{wang2017towards,miller2015discovering}.
However, being unreachable, these addresses are of no use to any other node.
At the same time, they increase network traffic~\cite{biryukov2014deanonymisation} and likely produce a high number of failed connection attempts.
Additionally, they potentially reduce the availability of reachable addresses, since new (often unreachable) addresses replace old ones in the node database when the address pool is full. 

From a security perspective, these addresses enable fingerprinting techniques, which allow for deanonymization attacks~\cite{biryukov2014deanonymisation,biryukov2015tor}.
Disabling their advertisement to outbound peers would effectively invalidate the few known deanonymization attacks targeting U nodes.

\subsection{Transaction Propagation Protocol}
\label{protocol}
In the following, we discuss and describe our new propagation protocol design.
The protocol explicitly leverages U nodes to improve resiliency against deanonymization.
Similarly to Dandelion~\cite{venkatakrishnan2017dandelion,fanti2018dandelion++}, we include two essential concepts in our design: proxied broadcast and transaction mixing. 
Proxied broadcast consists in delegating the diffusion of a new transaction to another node (called \textit{proxy}), allowing to hide the real origin of the transaction.
Mixing consists in sending to the proxy other new (proxied) transactions, received from other peers.
This makes it hard for the proxy to distinguish between transactions generated by the sender and others relayed by the sender, but generated by other nodes.

Given what said about U nodes in \S \ref{sect:unreachable}, we want to leverage their protected position in the network to conceal the origin of a transaction.
The core idea is to make R nodes, which are more susceptible to deanonymization, use U nodes as proxies for their transactions.
This way, such transactions will look as generated by their proxies instead of the actual source.
Additionally, we hinder proxies from distinguishing such transactions by mixing them with transactions from other nodes.

Before detailing the propagation protocol, we define our adversary model and motivate our design choices.

\paragraph{Network and Adversary Model}
\begin{figure}[tbp]
	\centerline{\includegraphics[width=0.7\columnwidth]{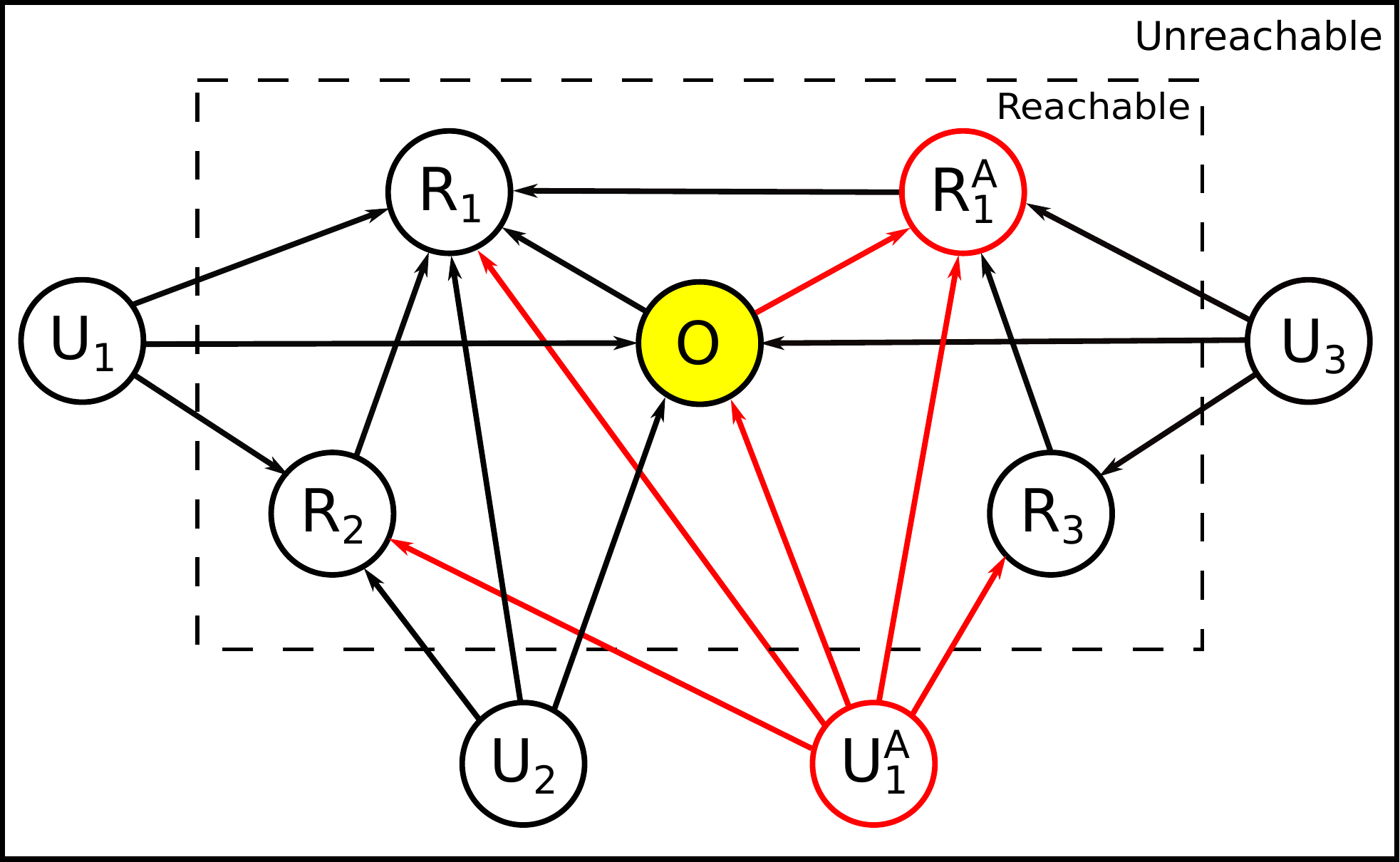}}
	\caption{Our view of the Bitcoin network: the origin O of a transaction is connected to R and U nodes. The adversary (colored in red) deploy both R and U nodes and connect to all reachable nodes.}
	\label{fig:netmodel}
\end{figure}
To describe our protocol, we model the Bitcoin network as in Figure \ref{fig:netmodel}.
We call $O$ the R node running the protocol and generating new transactions.
Other nodes are denoted by $R_i$, if reachable and $U_i$, if unreachable, for $i=1,2,...$.
The adversary $A$ aims at deanonymizing transactions generated by $O$ and can control various nodes, both reachable and unreachable, which we denote by $R^A_i$ and $U^A_i$, respectively.
$A$ can connect to all reachable nodes and also create multiple connections to the same node (including $O$).
Nonetheless, $A$ cannot directly connect to other U nodes.
To that respect, $A$ can only deploy multiple R nodes to increase the chance of having honest U nodes connecting to it.
Additionally, the adversary can create and transmit transactions, as well as relay or retain others received from its peers.

\paragraph{Design}
In our protocol, R nodes leverage U nodes as proxies and use transactions coming from other U peers for mixing.
Instead, U nodes use R nodes as proxies and mix new transactions with those coming from other R peers.

This scheme allows protecting both R and U nodes.
In fact, U nodes cannot distinguish between transactions generated by their R peers and those proxied by such peers but generated by other U nodes.
Similarly, U-generated transactions are indistinguishable to R nodes from those generated by other R nodes and proxied by their U peers.

However, a naive design could lead to easy deanonymization attacks, and also to an ineffective propagation of new transactions through the network.
Therefore, we need to define (1) which peers are used for proxying and (2) which transactions are used for mixing.

As for point (1), an R node can select one, all, or a subset of its U peers.
Note that an adversary can control a large subset of U peers of R nodes.
This increases her probability of being selected as the first proxy for many R-generated transactions, allowing an effective use of the first-spy estimator.
At the same time, if we send all transactions to a single proxy, it will be easy for this one to narrow down the set of transactions possibly generated by the sender.
As such, we first select a subset of peers to be used as proxy and pick a random one within this subset for every proxy operation.
We call this subset the \textit{proxy set}.
In order to distribute transactions among all nodes and minimize the risk of a proxy collecting all new transactions from a node, we change the proxy set at a certain rate. 
We call \textit{epoch} the time frame in which a proxy set is used.

As for point (2), we first need to identify which transactions are suitable for mixing.
Note that transactions received by an R node from other R peers following our protocol have already been diffused, making them unsuitable for mixing (since the adversary might already know them).
Similarly, transactions diffused by U peers might have already been received by the adversary.
On the other hand, it is easy to see that proxied transactions are the least likely to be known to the adversary, and thus best suite for mixing.

Therefore, we need to identify which transactions are being proxied and which are being diffused.
To do so, we mark proxied transactions and distinguish between two propagation phases: the \textit{proxying phase} and the \textit{diffusion phase}.
We call transactions in the proxying phase \textit{proxy transactions}.
When a new transaction is created is marked as proxying and sent to a node of the proxy set.
As for mixing, nodes use proxy transactions coming from their peers.
We call the set of proxy transactions used for mixing, the \textit{mixing set} of a node.
Transactions in the mixing set are relayed through the same path as newly-generated ones so as to make them indistinguishable from each other.

Ideally, we would like the mixing set to be as large as possible.
However, if we used all incoming proxy transactions, they would never be diffused.
Instead, we include only a fraction of such transactions in the mixing set, and diffuse the rest.
To do so, we need to decide which transactions to diffuse and which to relay.
A possible strategy is to select some peers in each epoch, and only use transactions coming from them.
However, if an adversary controls many of these peers and also the selected proxy, she could track most transactions in the mixing set of the target, leading to an easy deanonymization.
To avoid such a risk, we select proxy transactions from all of our peers and probabilistically include them in our mixing set.
In particular, for each proxy transaction, we keep proxying it with a certain probability $p$ and diffuse it otherwise.
This way, despite being able to track or inject proxy transactions for a specific node, an adversary cannot affect the number of honest transactions included in its mixing set.
A correct choice of $p$ will be fundamental for the effectiveness and efficiency of our protocol.

To further protect R nodes from adversaries controlling many inbound connections, 
we adopt the \textit{bucketing} strategy used in Bitcoin Core for managing addresses.
This mechanism is used to prevent an adversary from filling up the address database with malicious IPs,
and it is based on the assumption that the attacker only controls nodes from a limited address space~\cite{neudecker2015simulation}.
In particular, each bucket contains addresses from a different subnet.
Similarly, we make R nodes select proxies and transactions for the mixing set uniformly at random among peers from different buckets.

Finally, to cope with the risk of a transaction not being diffused, due to a DoS attack by a proxy or to an excessively long proxying phase, each node sets a timeout $t$ for every proxied transaction.
When $t$ expires, the node verifies if the transaction has been diffused by checking if the majority of outbound peers have advertised it back to us.
We choose to monitor outbound peers to minimize the risk of an adversary deceiving an R node by relaying proxied transactions from other adversary-controlled U peers.
The same rule is applied to both new and relayed transactions, so as to avoid deanonymization due to rebroadcast.
In the current protocol, in fact, a rebroadcast is only done by the source of the transaction, and can thus reveal its origin~\cite{koshy2014analysis}.
In our protocol, instead, rebroadcast applies to all proxied transactions, thus leaking no new information.

\paragraph{Protocol Rules}
To detail the propagation rules of our protocol, we first define the \textit{proxy} operation on a transaction $tx$ as follows:
\begin{algorithm}
\SetAlgoLined
Pick a random peer $P$ from the proxy set\;
Send $tx$ to $P$ and set a timeout $t$\;
When $t$ expires:\\
\eIf {\text{The majority of outbound peers advertised} $tx$}
{Return}
{Repeat}
\caption{Proxy($tx$)}
\end{algorithm}

Next, we define the propagation rules for R nodes:
\begin{algorithm}
\SetAlgoLined
Divide time into epochs\;
\If{\text{New epoch begins}}
{Select subset $S$ from U peers uniformly at random from different buckets\;
 Set $S$ as the proxy set}
  
\If{\text{Create new transaction $tx$}}
{Mark $tx$ as $proxying$\;
 Run $proxy(tx)$} 
\If{\text{Receive a $proxying$ transaction $tx_m$ from a U peer}}
{with probability $p$, execute $proxy(tx_m)$\;
 otherwise, $diffuse(tx)$}

\caption{R Propagation Rules}
\end{algorithm}

U nodes follow the same rules, except they use R peers instead of U peers and do not use buckets.

\section{Discussion}
\subsection{Limitations}
Our protocol requires R nodes to have U peers connected to them.
However, newly-joined R nodes usually have to wait some time to have other peers connect to them.
We address this limitation by having new R nodes using the diffusion protocol until they have a sufficient number of U peers.
Additionally, to prevent an adversary from taking advantage from this situation (by filling up all inbound slots), we also adopt the bucketing strategy.
Specifically, we make R nodes use our protocol only when enough U peers from different buckets are connected.

\subsection{Propagation and anonymity}
To better understand our protocol it is useful to depict the propagation pattern of a transaction.

Let us consider an R node $O$ generating a transaction $tx$.
The following sequence of events happens:
\begin{enumerate}
	\item R selects a proxy $P$ among its proxy set, mark $tx$ as $proxying$ and sends it to $P$;
	\item $P$ receives $tx$ and proxy it with probability $p$, or diffuses it otherwise;
	\item If proxying $tx$, $P$ selects a node $R$ from its proxy set $S$ and sends it $tx$;
\end{enumerate}
Proxying transactions are relayed through a sequence of R and U nodes until it gets diffused.
Diffusion can happen at any step, except for the first one.
Propagation from an U node follows a similar pattern.

A major risk of proxied broadcast is that a transaction might take too long to diffuse, or not be diffused at all.
As for diffusion time, we can statistically guarantee to diffuse every transaction within a reasonable time.
Since at every hop, the transaction $tx$ is diffused with probability $p$, it is possible to tune this value to obtain a target number of hops through which $tx$ is proxied on average.
The use of timeouts allows dealing with a transaction not being diffused.

With respect to anonymity, our protocol is designed to be resistant against a first-spy estimator.
This type of adversary connects to all R nodes and links each transaction to the first node from which it has been received.
As demonstrated in \cite{fanti2017anonymity}, this strategy is very effective with the current propagation model.
However, the changes introduced by our protocol make it very unlikely for a node to first receive a transaction from its source.
On the contrary, most of the times, transactions will be received by a node different from the origin, thanks to proxying.
Furthermore, each transaction is mixed with many others generated by nodes in the proxying path, which are indistinguishable from each other to the receiving node.
This means that any claim about the origin of a transaction can be easily denied.

Note that our protocol is designed to resist against very powerful adversaries controlling several nodes and maintaining multiple connections to all reachable nodes.
The adversary can combine information from all of its nodes and coordinate them to influence or track the mixing set of a target node.
However, we showed in the previous section how such an adversary has limited capabilities to affect the security of the protocol.

\subsection{Ephimerality of U nodes}
A possible issue in our design is the short time of connection of many U nodes.
In fact, while R nodes are relatively stable \cite{statoshi-peers}, U nodes often experience very short-lived connections \cite{wang2017towards}.
This behavior might affect the efficiency of the protocol.
However, the timeout mechanism is also meant to deal with this kind of problems and can be fine-tuned independently by each node, depending on the experienced churn.

Moreover, the presence of short-lived proxy nodes, if properly exploited, might serve as an added value to the anonymity level of our protocol, as it makes it harder to track back a transaction to its origin.

\subsection{NAT adoption}
Another potential limitation of our solution is that it is based on the unreachability of NATted nodes.
However, if IPv6 gets adopted by the majority of nodes, it is possible that NATs will cease to be used. 
The introduction of IPv6, in fact, was mainly intended to deal with the IPv4 address exhaustion problem and remove the need for NATting~\cite{rfc2007ip6}.

Although growing, the adoption rate of IPv6 seems to be variable \cite{mccarthy2018ipv6} and not uniform worldwide, with statistics strongly dependent on the adopted metrics ~\cite{czyz2014measuring}.
As for the Bitcoin network, Neudecker et al.~\cite{neudecker2019characterization} showed that, unlike IPv4, IPv6 connections have not grown over the past two years.

Either way, most optimistic estimates predict a complete adoption within 7-8 years~\cite{howard2019ipv6}.
In this perspective, our protocol should be considered as a medium-term solution, likely able to work for the next decade.

\section{Related Work}
\label{sec:related}
\subsection{Unreachable nodes}
Unreachable nodes have been extensively studied by Wang et al.~\cite{wang2017towards}.
In order to perform their analyses, they deployed around 100 nodes, through which they collected information on more than 100 K unreachable peers, which generated more than 2 M transactions.
Their findings show that most connections last for less than 60 seconds, while, at the same time, most transaction propagations are sent over long-lived connections (more than 100 seconds), showing a high degree of centrality.
Finally, they show a method to deanonymize transactions coming from unreachable peers, with the help of an external listener node.
Their results show that unreachable nodes are also susceptible to the first-spy estimator attack.
Note that our protocol makes this attack much less effective, since new transactions are proxied to a single reachable node, reducing the probability that the attacker receives the transaction.
At the same time, new transactions are mixed with transactions proxied from its peers, thus reducing the accuracy of the attack.

Other deanonymization attacks also target U nodes globally by means of fingerprinting techniques.
Biryukov et al. \cite{biryukov2014deanonymisation} make use of ADDR messages to uniquely identify U nodes by the set of their peers.
Their technique allows linking multiple transactions created by the same node over a single session.
In this paper, we proposed a change in the address advertisement by U nodes that would make this attack ineffective.

In \cite{biryukov2015tor}, Biryukov et al. devise a technique to deanonymize U nodes connecting via Tor, even through multiple sessions. 
However, their attack is specific to Tor users.

Finally, Mastan et al. \cite{mastan2018new} exploit block requests patterns to identify U nodes over consecutive sessions.
However, their technique only allows linking sessions and thus needs to be used in conjunction with other deanonymization techniques.

\subsection{Transaction Propagation Anonymity}
Anonymity properties of transaction propagation have been extensively studied in research.

Koshy et al. \cite{koshy2014analysis} are among the first ones to show a practical deanonymization technique for Bitcoin, based on transaction propagation analysis.
They show that anomalies in the propagation can be exploited to identify the source of transactions.

Neudecker et al.~\cite{neudecker2017could} combine observations of the message propagation with Bitcoin address clustering techniques. 
However, their results show that for the vast majority of users this information does not facilitate deanonymization.

In \cite{fanti2017anonymity}, Fanti et al. thoroughly analyze the anonymity properties of both the former and the actual Bitcoin propagation protocols (Trickle and Diffusion).
They theoretically prove that both protocols offer poor anonymity on networks with a regular-tree topology.
They also identify the symmetry of current spreading protocols as the main characteristic that allows deanonymization attacks.
An alternative protocol, called Dandelion, is then proposed in \cite{venkatakrishnan2017dandelion,fanti2018dandelion++} that specifically addresses this issue.
Dandelion breaks the symmetry by proxying the broadcast of new transactions through a network-wide circuit of nodes.
Furthermore, they increase anonymity by mixing new transactions over the same path.
However, their protocol is somewhat hard to implement and only applies to R nodes. 
Our protocol, instead, involves both R and U nodes and does not seem to show any difficulties of implementation.

\section{Conclusion and Future Work}
\label{sec:conclusion}

In this paper, we showed how unreachable nodes are often overlooked despite being a relevant part of the Bitcoin network.
We also showed how increasing their connectivity and participation can be beneficial for the robustness and efficiency of the whole network, without introducing overhead.
Moreover, we leveraged their peculiar position in the network to design a new transaction propagation protocol to protect all nodes from deanonymization attacks.
We analyzed the security of our proposal against powerful adversaries and discussed possible limitations of our approach.

Future work includes implementing our changes and evaluating their effectiveness through experiments.
A more formal analysis is also needed to provide better anonymity guarantees.
Finally, an analysis of the current participation degree of unreachable nodes in the transaction propagation might shed light on their relevance in the Bitcoin network.



%
\printbibliography

\end{document}